\documentclass[preprint]{revtex4}
\usepackage{graphicx}

\begin{document}
\title{Quantum Phase Transition of Spin-2 Cold Bosons in an Optical Lattice}

\author{Jing-Min Hou}
\email{jmhou@eyou.com}
\affiliation{Theoretical Physics Division,
Nankai Institute of Mathematics, Nankai University, Tianjin,
300071, China} \affiliation{Liuhui Center for Applied Mathematics,
Tianjin, 300071, China}
\author{Mo-Lin Ge}
\affiliation{Theoretical Physics Division, Nankai Institute of
Mathematics, Nankai University, Tianjin, 300071, China}
\affiliation{Liuhui Center for Applied Mathematics, Tianjin,
300071, China}

\begin{abstract}
   The Bose-Hubbard Hamiltonian of spin-2 cold bosons with repulsive interaction in an optical
   lattice is proposed. After neglecting the hopping term, the site-independent Hamiltonian and its energy eigenvalues and
   eigenstates
   are obtained. We consider the hopping term as a perturbation to do the calculations in second order and draw the phase
   diagrams for different cases. The phase diagrams show that  there
   is a phase transition from Mott insulator with integer number bosons
   to superfluid when the ratio
   $c_0/t$ ($c_0$ is the spin-independent on-site interaction and $t$ the hopping matrix element
    between adjacent lattice sites)
   is decreased to a critical value and that there is different phase boundary between superfluid and Mott
   insulator phase for different Zeeman level component in some
   ground states. We find that the position of  phase
   boundary for different Zeeman level component is related to
   its average population in the Mott ground state.\\
   PACS number(s):03.75.Kk, 03.75.Lm, 03.75.Mn, 32.80.Pj
\end{abstract}
\maketitle
\section{Introduction}
During recent years, the superfluid-Mott insulator transition
becomes a significant topic of atomic physics, which has been
extensively studied in the context of $^4$He absorbed in the
porous media\cite{fisher}. The detailed study of superfluid-Mott
insulator transition becomes possible when quantum phase
transition from a superfluid  to a Mott insulating ground state
was observed in a Bose-Einstein condensate stored in a three
dimensional optical lattice potential by Greiner et
al.\cite{greiner}. The optical lattices used to confine cold atoms
are induced by the ac Stark effect of interfering laser beams. The
dynamics of the bosonic atoms in optical lattices realizes a
Bose-Hubbard model\cite{jaksch}, which predicts phase transition
from a superfluid(SF) to Mott insulator(MI) at low temperature
with increasing the ratio of the on site interaction $c_0$ to the
hopping matrix element $t$.

Besides many experimental efforts made to realize superfluid-Mott
insulator transition, a large number of theoretical literatures
appeared\cite{fisher,jaksch,oosten,chen,demler,tsuchiya}.
Ref.\cite{oosten} made an appropriate mean-field approximation to
the Hamiltonian of spinless or polarized bosons in an optical
lattice to describe the zero-temperature phase transition from the
superfluid to the Mott-insulating phase analytically and
calculated the phase diagram. In this paper a first-order
approximation to the dispersion of the density fluctuations shows
that the system indeed goes from a gapped to a gapless phase.

Since trapping a Bose condensate by purely optical means is
realized several years ago\cite{stamper-kurn}, which liberates the
internal degrees of freedom of spinor Bosons that are frozen in a
magnetic trap, much attention is attracted on the study of spinor
Bose condensate, which has multiple components. Ho\cite{ho} and
Ohmi and Machida\cite{ohmi} have researched a spin-1 BEC and
obtained the general theoretical frameworks respectively.
Ciobanu,Yip and Ho\cite{ciobanu} have studied an $F=2$ spinor bose
condenate and drawn the phase diagrams. Koashi and Ueda have
obtained the exact eigenstates of spin-1 and spin-2 vertical BEC
and discussed their magnetic response\cite{koashi,koashi2}.

The next natural step after the detailed research of spinor bose
condensates with free internal freedom is the study of the spinor
multi-component BEC in a optical lattice---a new background.
Demler and Zhou have studied spin-1 Bose atoms in an optical
lattice  and obtained  several unique properties\cite{demler}.
Ref.\cite{tsuchiya} studied the spin-1 bosons interacting
antiferromagnetically in an optical lattice with the mean-field
approximation method and obtained the phase diagram showing a
transition from Mott insulator to superfluid, which is signed by
the appearing of the superfluid order parameter i.e. the density
fluctuations. But the spin-2 cold Bose atoms in an optical lattice
have not been researched as yet, which is the principal task of
this paper.

In this paper, we study the transition from Mott insulator to
superfluid of spin-2 Bose atoms with repulsive interaction  in
terms of mean-field approximation method. First, ignoring the
hopping term of the Hamiltonian, we get the eigenstates and
eigenvalues and find the ground states of the system for different
cases. Then, we consider the hopping term as a perturbation to do
calculations in second order and draw the phase diagrams.
\section{The model}
We consider bosons with hyperfine spin $F=2$, such as $^{23}{\rm
Na}$,$ ^{87}{\rm Rb}$ or $^{85}{\rm Rb}$, in an optical lattice.
Adding the external periodic potential term $V({\bf
r})=V_0(\sin^2kx+\sin^2ky+\sin^2kz)$, where $V_0$ is a tunable
amplitude and $k$ the wave vector of the laser light, to the
Hamiltonian of spin-2 bosons\cite{ciobanu}, we get the Hamiltonian
of spin-2 bosons in an optical lattice in second quantized form as
follows
\begin{eqnarray}
H=\int d{\bf
r}[\frac{\hbar}{2m}\nabla\Psi_\alpha^\dag\cdot\nabla\Psi_\alpha+V({\bf
r}) \Psi_\alpha^\dag\Psi_\alpha
-\mu\Psi_\alpha^\dag\Psi_\alpha+\nonumber
\frac{\bar{c}_0}{2}\Psi_\alpha^\dag\Psi_\beta^\dag\Psi_\beta\Psi_\alpha
\\+\frac{\bar{c}_1}{2}\sum_i(\Psi_\alpha^\dag(F_i)_{\alpha\beta}\Psi_\beta)^2
+\bar{c}_2\Psi_\alpha^\dag\Psi_{\alpha^\prime}^\dag\langle
2\alpha;2\alpha^\prime|00\rangle\langle00|2\beta;2\beta^\prime\rangle\Psi_\beta\Psi_{\beta^\prime}]
\label{a}
\end{eqnarray}
where $m$ is the atomic mass, $\Psi_{+2},...,\Psi_{-2}$ are the
 five-component field operators corresponding to the sublevels
$m_F=+2,...,-2$ of the hyperfine state $F=2$, $\mu$ is the
chemical potential, $\bar{c_0}$, $\bar{c_1}$ and $\bar{c_2}$ are
related to s-wave scattering length $a_0$,$a_2$ and $a_4$ of the
two colliding bosons with total angular momenta $0, 2$ and $4$ by
$\bar{c_0}=4\pi \hbar^2(3a_4+4a_2)/7m$, $\bar{c_1}=4\pi
\hbar^2(a_4-a_2)/7m$ and $\bar{c_2}=4\pi
\hbar^2(3a_4-10a_2+7a_0)/7m$, $<2\alpha;2\alpha^\prime|00>$ and
$<00|2\beta;2\beta^\prime>$ are Clebsch-Gordan coefficients.  $
F_\alpha(\alpha=x,y,z)$ are $5\times5$ spin matrices satisfying
the commutation relation $[F_\alpha,
F_\beta]=\varepsilon_{\alpha\beta\gamma}F_\gamma$.

For a single atom in the periodic potential, the energy
eigenstates are Bloch states. In the tight-binding limit, we can
superpose the Bloch states to get a set of Wannier functions,
which are localized on an individual lattice site. Expanding the
field operators in the Wannier basis and keeping only the lowest
vibrational states, $\Psi_\alpha=\sum_i b_{i\alpha} w({\bf r}-{\bf
r}_i)$, Eq.(\ref{a}) reduces to the Bose-Hubbard Hamiltonian
\begin{equation}
H=-t\sum_{<i,j>}b_i^\dag b_j
-\mu\sum_i\hat{n}_i+\frac{c_0}{2}\sum_i\hat{n}_i(\hat{n}_i-1)+\frac{c_1}{2}\sum_i(\hat{\bf{F}}_i^2-6\hat{n}_i)
+\frac{2c_2}{5}\sum_i\hat{S}_{i+}\hat{S}_{i-}
\label{b}
\end{equation}
where $\hat{\bf F}_i=b_{i\alpha}^\dag {\bf
F}_{\alpha\beta}b_{i\beta}$, and $\hat{n}_i=\sum_\alpha
b_{i\alpha}^\dag b_{i\alpha}$. $t=-\int d{\bf r}w_i^*({\bf
r})(-\hbar^2\nabla^2/2m+V({\bf r}))w_j({\bf r})$ is the hopping
matrix element between adjacent sites $i$ and $j$. $c_i$ is
on-site inter-atom interaction defined by $c_i=\bar{c}_i\int d{\bf
r}|w_i({\bf r})|^4$, where the Hubbard approximation has been used
to approximate the multi-center integral as a single-center one.
$\hat{S}_{i+}=\hat{S}_{i-}^\dag=(b_{i0}^\dag)^2/2-b_{i1}^\dag
b_{i-1}^\dag+b_{i2}^\dag b_{i-2}^\dag$, when applied to the
vacuum, creates a spin-singlet ``pairs''.  The properties of the
operators will be discussed in detail in the following section.
The fourth and fifth terms in the equation are adding terms
comparing with the Bose-Hubbard Hamiltonian of  spinless or
polarized bosons\cite{jaksch}. By contract with that of spin-1
bosons\cite{tsuchiya}, the fifth term is an adding term and
$6\hat{n}_i$ in the fourth term replaces $2\hat{n}_i$.

In the limit $c_0/t\longrightarrow \infty$, the hopping term in
the Hamiltonian can be neglected, so the Hamiltonian  is reduced
to a diagonal matrix with respect to sites. Omitting the site
index, the singe-site Hamiltonian is
\begin{equation}
h_0=-\mu\hat{n}+\frac{c_0}{2}\hat{n}(\hat{n}-1)+\frac{c_1}{2}(\hat{\bf
F}^2-6\hat{n})+\frac{2c_2}{5}\hat{S}_+\hat{S}_- \label{c}
\end{equation}
whose eigenstates and eigenvalues  will be discussed in following
section.

To study the quantum transition, we use the mean-field
approximation to the hopping term and consider it as a
perturbation. Here we introduce the superfluid order parameter
$\phi_\alpha=<b_{i\alpha}>=\sqrt{n_{sf}}\zeta_\alpha$, where
$n_{sf}$ is the superfluid density and $\zeta_\alpha$ is a
normalized spinor $\zeta_\alpha^*\zeta_\alpha=1$.  The hopping
term is decoupled as $ b_{i\alpha}^\dag
b_{j\alpha}\approx(\phi_\alpha b_{i\alpha}^\dag+\phi_\alpha^*
b_{j\alpha})-\phi_\alpha^*\phi_\alpha$. So the total hopping term
becomes the product of a site-independent term and the total
number of the sites. We can consider only a single site because
the Hamiltonian of  every site is identical in the homogenous
case. In the single site, the resulting mean-field version of the
hopping Hamiltonian can be written as
\begin{equation}
h_1=zt(\phi_\alpha b_\alpha^\dag+\phi_\alpha^*
b_\alpha-\phi_\alpha^*\phi_\alpha) \label{d}
\end{equation}
where $z$ is the number of the nearest-neighbor sites and the
site-index is omitted. So the Hamiltonian of a single site is
\begin{equation}
h=h_0+h_1
\end{equation}
When the ratio $c_0/t$ is very large, we can consider $h_1$ as a
perturbative term.
\section{The energy eigenvalues and the ground states}
Before  perturbative calculations we should get the eigenvalues
and eigenstates of Eq.(\ref c), i.e.,  to sovle the equation
\begin{equation}
h_0\psi=\varepsilon^{(0)}\psi
\end{equation}
 In Eq.(\ref c), the operaters $\hat{S}_+$ and $\hat{S}_-$ satisfy
the $SU(1,1)$ commutation relations together with
$\hat{S}_z\equiv(2\hat{n}+5)/4$, namely\cite{koashi,koashi2},
\begin{equation}
[\hat{S}_z,\hat{S}_\pm]=\pm\hat{S}_\pm,\ \ \
[\hat{S}_+,\hat{S}_-]=-2\hat{S}_z \label{e}
\end{equation}
and the Casimir operator $\hat{\textbf{S}}^2$  commuting  with
$\hat{S}_\pm$ and $\hat{S}_z$ reads
\begin{equation}
\hat{\textbf{S}}^2\equiv-\hat{S}_+\hat{S}_-+\hat{S}_z^2-\hat{S}_z
\label{f}
\end{equation}
The eigenvalues of the mutual eigenstates for $\hat{\textbf{S}}^2$
and $\hat{S}_z$ are $\{S(S-1),S_z\}$ with $S=(2n_0+5)/4\ \
(n_0=0,1,2,...)$ and $S_z=S+n_s\ \ (n_s=0,1,2,...)$, which
guarantees that
$\hat{S}_+\hat{S}_-=\hat{S}_z^2-\hat{S}_z-\hat{S}^2$ is positive
semidefinite. The new quantum numbers $n_s$ and $n_0$ are
introduced and the operator $S_+$ raises $n_s$ by one while the
relation $n=2n_s+n_0$ holds, where $n$ is the total number of
bosons in a single site.

The operators $\hat{S}_\pm$ commute with the spin operator
$\hat{\textbf{F}}$ and the magnetic quantum number operator
$\hat{F}_z$, So the energy eigenstates can be classified according
to quantum numbers $n_0$ and $n_s$, total spin $F$, and magnetic
quantum number $F_z$. Hence, the eigenstates $\psi$ are denoted as
$|n_0,n_s,F,F_z;\lambda>$ where $\lambda$ labels orthonormal
degenerate states. The energy eigenvalue  is given by
\begin{equation}
\varepsilon^{(0)}=-\mu
n+\frac{c_0}{2}n(n-1)+\frac{c_1}{2}[F(F+1)-6n]+\frac{2c_2}{5}n_s(n-n_s+\frac{3}{2})
\label{energy}
\end{equation}
where the relation $n=2n_s+n_0$ is used. So Mott states can be
expressed as $\prod_i|n_0,n_s,F,F_z;\lambda>_i$, where $i$ is
lattice site index. For homogenous case, the zeroth order total
energy is $E^{(0)}=\sum_i \varepsilon^{(0)}=N_l\varepsilon^{(0)}$,
where $N_l$ is the number of lattice sites.

From Ref.\cite{koashi2} we know that the energy eigenstates
$|n_0,n_s,F,F_z;\lambda>$ can be represented as
\begin{equation}
(\hat{F}_-)^{\Delta
F}(\hat{A}_0^{(2)\dag})^{n_{20}}\hat{P}_{(n_s=0)}(b_2^\dag)^{n_{12}}
(\hat{A}_2^{(2)\dag})^{n_{22}}(\hat{A}_0^{(3)\dag})^{n_{30}}(\hat{A}_3^{(3)\dag})^{n_{33}}|vac>
\end{equation}
where
\begin{eqnarray}
\hat{A}_0^{(2)\dag}&=&\frac{1}{\sqrt{10}}[(b_0^\dag)^2-2b_1^\dag
b_{-1}^\dag+2b_2^\dag b_{-2}^\dag]\\
\hat{A}_2^{(2)\dag}&=&\frac{1}{\sqrt{14}}[2\sqrt{2}b_2^\dag
b_0^\dag-\sqrt{3}(b_1^\dag)^2]\\
\hat{A}_0^{(3)\dag}&=&\frac{1}{\sqrt{210}}[\sqrt{2}(b_0^\dag)^3-3\sqrt{2}b_1^\dag
b_0^\dag
b_{-1}^\dag+3\sqrt{3}(b_1^\dag)^2b_{-2}^\dag+3\sqrt{3}b_2^\dag
(b_{-1})^2-6\sqrt{2}b_2^\dag b_0^\dag b_{-2}^\dag]\\
\hat{A}_3^{(3)\dag} &=& \frac{1}{20}[(b_1^\dag)^3-\sqrt{6}b_2^\dag
b_1^\dag b_0^\dag+2(b_2^\dag)^2b_{-1}^\dag]
\end{eqnarray}
$P_{(n_s=0)}$ is the projection onto the subspace with $n_s=0$;
$n_{12},n_{20},n_{22},n_{30}=0,1,2,...,\infty$, $n_{33}=0,1$, and
$\Delta F=0,1,...,2F$ that are related to $\{n_0,n_s,F,F_z\}$ as
\begin{eqnarray}
n_0&=&n_{12}+2n_{22}+3n_{30}+3n_{33}\\
n_s&=&n_{20} \\
F&=&2n_{12}+2n_{22}+3n_{33}\\
F_z&=&F-\Delta F
\end{eqnarray}
  To find the ground
states of Eq.(\ref{c}), we minimize its energy eigenvalue, i.e.,
Eq.(\ref{energy}). We note that there isn't magnetic quantum
number $F_z$ in Eq.(\ref{energy}), which means the eigenstates
with different magnetic quantum number but the same other quantum
numbers are degenerate. For simplicity, in this paper we only
investigate the states with the highest magnetic quantum number,
i.e. the states with $F_z=F$. when $n=1$, it is obvious that the
ground state is $|1,0,2,2;\lambda>$, which means  only one single
atom with the highest magnetic quantum number for the state. For
$n\geq 2$, it is more complicated, which will be discussed in
detail as follow:
\begin{enumerate}
\item{$c_1>0,\ \ c_2>0$. In this case, if $n_s=0$ and $F=0$, the
energy eigenvalue has its minimum. But that is not always the case
because there are some forbidden values\cite{koashi,koashi2},
i.e., $F=1,2,5,2n_0-1$ are not allowed when $n_0=3k(k\in Z)$, and
$F=0,1,3,2n_0-1$ are forbidden when $n_0=3k\pm 1(k\in Z)$,
according to which, it is classified into three cases.
  \begin{enumerate}
  \item{For $n=3k(k\in Z)$, $|n,0,0,0;\lambda>$ with $n_s=0$
  and $F=0$ simultaneously, which is not within forbidden values,
  is the ground state}
  \item{For $n=3k-1(k\in Z)$, the state with $n_s=0$ and $F=0$
  simultaneously is not allowed. If $n_s=0$, the lowest allowed
  value of total spin $F$ is $2$.  $F=0$ is not forbidden
  when $n_0$ is $3k(k\in Z)$. $n_s=1$, due to the relation
  $n=2n_s+n_0$, is the lowest value satisfying the condition. So
  the state with $n_s=1$ and $F=0$ is a possible ground state.
  There is competition between the third term and the fourth
  term in Eq.(\ref{energy}),i.e.,  competition between
  contributions of total spin and singlet ``pairs'' to eigenenergy.
  Comparing both eigenenergies for the two cases $F=0,n_s=1$ and
  $F=2,n_s=0$, we get the ground state, (i)$|n-2,1,0,0;\lambda>$
  for $c_1>c_2(2n+1)/15$ and (ii) $|n,0,2,2;\lambda>$ for
  $c_1<c_2(2n+1)/15$.}
   \item{For $n=3k+1(k\in Z)$, it is similar to last case that there is
   competition between contributions of total spin and the singlet
   ``pair''
   to eigenenergy in this case. But there is a difference between the two
   cases in that if $F=0$, $n_s$ is at least $2$. Therefore, the ground state is
   (i)$|n-4,0,2,2;\lambda>$ for $c_1>2c_2(2n-1)/15$; (ii)
   $|n,0,2,2;\lambda>$ for $c_1<2c_2(2n-1)/15$, which is obtained
   by comparing the eigenenergy of these states.}
 \end{enumerate}}
\item{$c_1>0,\ \ c_2<0$. In this case, the eigenenergy is the
lowest when $F$ is at the lowest value and $n_s$ at the highest
value. There is a state with $F=0$ and $n_s=n/2$ when $n$ is even,
but there is not when $n$ is odd. So, when $n$ is different, there
are two cases as follow:
  \begin{enumerate}
  \item{When $n$ is even, the ground state is $|n,n/2,0,0;\lambda>$}
  \item{When $n$ is odd, $n_s$ has the highest value $(n-1)/2$. But $F$ is not zero  when $n_s=(n-1)/2$.
  Alternatively, there is another case that $F=0$ and $n_s=(n-3)/2$. In the two cases, the state that
  has lower eigenenergy is the ground state. Hence, the ground state is (i) $|1,(n-1)/2,2,2;\lambda>$
   for $c_1<7|c_2|/15$; (ii) $|3,(n-3)/2,0,0;\lambda>$ for $c_1>7|c_2|/15$.}
  \end{enumerate}}
\item{For $c_1<0,\ \ c_2>0$, the eigenstate with $F=2n$ and $n_s=0$, which has the lowest eigenenergy,
is the ground state. So the ground state is
$|n,0,2n,2n;\lambda>$.}
\item{For $c_1<0,\ \ c_2<0$, to get the ground state, we minimize Eq.(\ref{energy}). If we skip over the fact
for the moment that the singlet ``pair'' number $n_s$ is an
integer, then the condition to minimize the energy function is
given by}
\begin{equation}
n_s=\frac{10c_1(4n+1)-c_2(2n+3)}{80c_1-4c_2}
\end{equation}
because of the fact that the singlet ``pair'' number must be an
integer, we write $n_s$ in terms of the closet integer number
$n_s^0$ and the decimal part, i.e. $n_s=n_s^0+\alpha$, where the
number $\alpha$ satisfy $-1/2<\alpha<1/2$, which can be rewritten
as,
  \begin{equation}
   n_s^0-\frac{1}{2}<\frac{10c_1(4n+1)-c_2(2n+3)}{80c_1-4c_2}<n_s^0+\frac{1}{2}
   \end{equation}
\begin{enumerate}
  \item{Because the singlet ``pair'' number isn't negative, for $n_s^0<0$, $n_s$ must be zero. Hence,
  in this case, the ground state is $|n,0,2n,2n;\lambda>$.}
  \item{$0<n_s^0<\frac{n}{2}$. In this case, the eigenenergy is lower when the singlet ``pair'' number is $n_s^0$
  than any other integer. So the ground state is $|n-2n_s^0,n_s,2n-4n_s^0,2n-4n_s^0;\lambda>$.}
  \item{For $n_s^0>\frac{n}{2}$, the singlet ``pair'' number takes the highest value as it can. With different total
  bosons number, there are two cases as follow:
  (i)when $n$ is even, the ground state is $|0,n/2,0,0;\lambda>$;
   (ii)when $n$ is odd, the ground state is $|1,n/2,2,2;\lambda>$.}

  \end{enumerate}
\end{enumerate}
\section{The superfluid-Mott insulator transition}
 In this section, we consider the hopping
term and calculate the first and second order corrections to the
ground energy.
 Although the
eigenstates, among whose quantum numbers only magnetic quantum
number is different, are degenerate, we can do perturbative
calculations regarding the degenerate states as nondegenerate ones
for the off-diagonal hopping matrix elements between  states with
the same Boson number $n$ are zero, i.e.,
$<n_0,n_s,F,F_z;\lambda|h_1|n_0,n_s,F,F_z^\prime;\lambda>=0$, when
$F_z\neq F_z^\prime$ . Therefore, the first and second order
corrections to  ground energy are expressed as follow
\begin{equation}
\varepsilon_g^{(1)}=<g|h_1|g>=zt\sum_\alpha
\phi_\alpha^*\phi_\alpha\ \ \ \alpha=-2,...,2
\end{equation}
\begin{equation}
\varepsilon_g^{(2)}=\sum_{n\neq
g}\frac{|<g|h_1|n>|^2}{\varepsilon_g^{(0)}-\varepsilon_n^{(0)}}=\sum_{n\neq
g}\sum_\alpha\frac{z^2t^2|<g|b_\alpha+b_\alpha^\dag|n>|^2\phi_\alpha^*\phi_\alpha}{\varepsilon_g^{(0)}-\varepsilon_n^{(0)}}\
\ \ \alpha=-2,...,2
\end{equation}
The ground energy modified by adding the first and second order
corrections becomes
\begin{equation}
\varepsilon_g=\varepsilon_g^{(0)}+\varepsilon_g^{(1)}+\varepsilon_g^{(2)}=\varepsilon_g^{(0)}+zt\sum_\alpha
A_\alpha(n,\bar{\mu},\bar{c}_0,\bar{c}_1,\bar{c}_2)\phi_\alpha^*\phi_\alpha\
\ \ \alpha=-2,...,2 \label{ge}
\end{equation}
where $A_\alpha(n,\bar{\mu},\bar{c}_0,\bar{c}_1,\bar{c}_2)$ is
related to the first and second order corrections of the spin
component with magnetic quantum number $\alpha$ to zero-order
ground energy. It depends on  $n$, $\bar{\mu}$,
$\bar{c}_0$,$\bar{c}_1$ and $\bar{c}_2$, where $\bar{\mu}=\mu/zt,
\bar{c}_0=c_0/zt,\bar{c}_1=c_1/zt,\bar{c}_2=c_2/zt$.  Minimizing
the ground energy function (\ref{ge}) modified in second order, we
find that $\phi_\alpha=0$ when
$A_\alpha(n,\bar{\mu},\bar{c}_0,\bar{c}_1,\bar{c}_2)>0$ and that
$\phi_\alpha\neq 0$ when
$A_\alpha(n,\bar{\mu},\bar{c}_0,\bar{c}_1,\bar{c}_2)<0$. This
means that $A_\alpha(n,\bar{\mu},\bar{c}_0,\bar{c}_1,\bar{c}_2)=0$
signifies the boundary between the superfluid and the Mott
insulator phases of the spin component with magnetic quantum
number $\alpha$.

\begin{figure}[ht]
 \includegraphics[width=0.8\columnwidth]{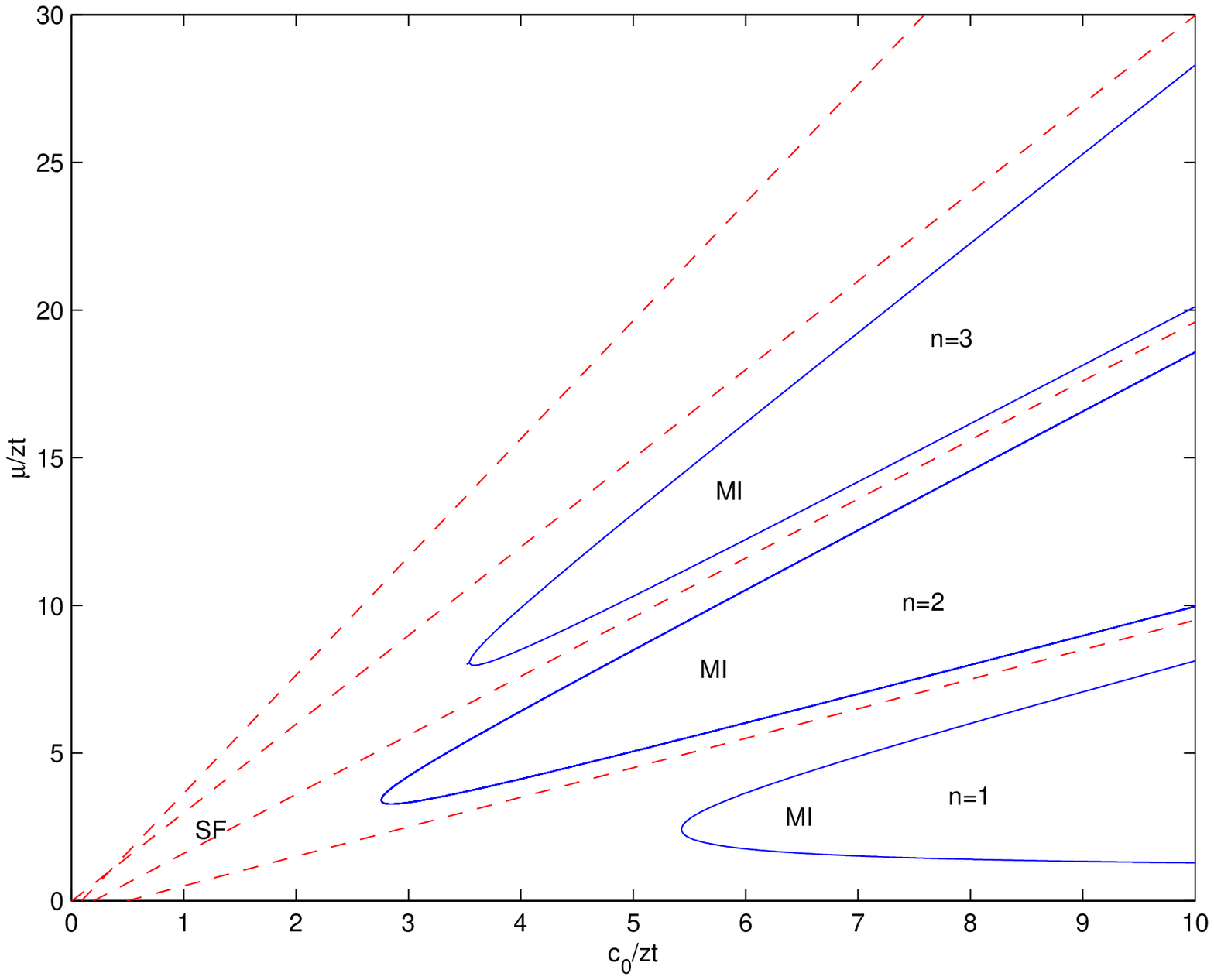}
 \caption{Phase diagram
of Bose-Hubbard Hamiltonian  obtained from second-order
perturbation theory with solid lines for $c_1=0.1zt$ and
$c_2=0.1zt$. Here $c_0/zt$ and $\mu/zt$ are dimensionless. The
dashed lines indicate the zeroth order phase diagram. In the
diagram, SF and MI denote superfluid phase and Mott insulator
phase respectively.}\label{1}
\end{figure}

\begin{figure}[ht]
\includegraphics[width=0.8\columnwidth]{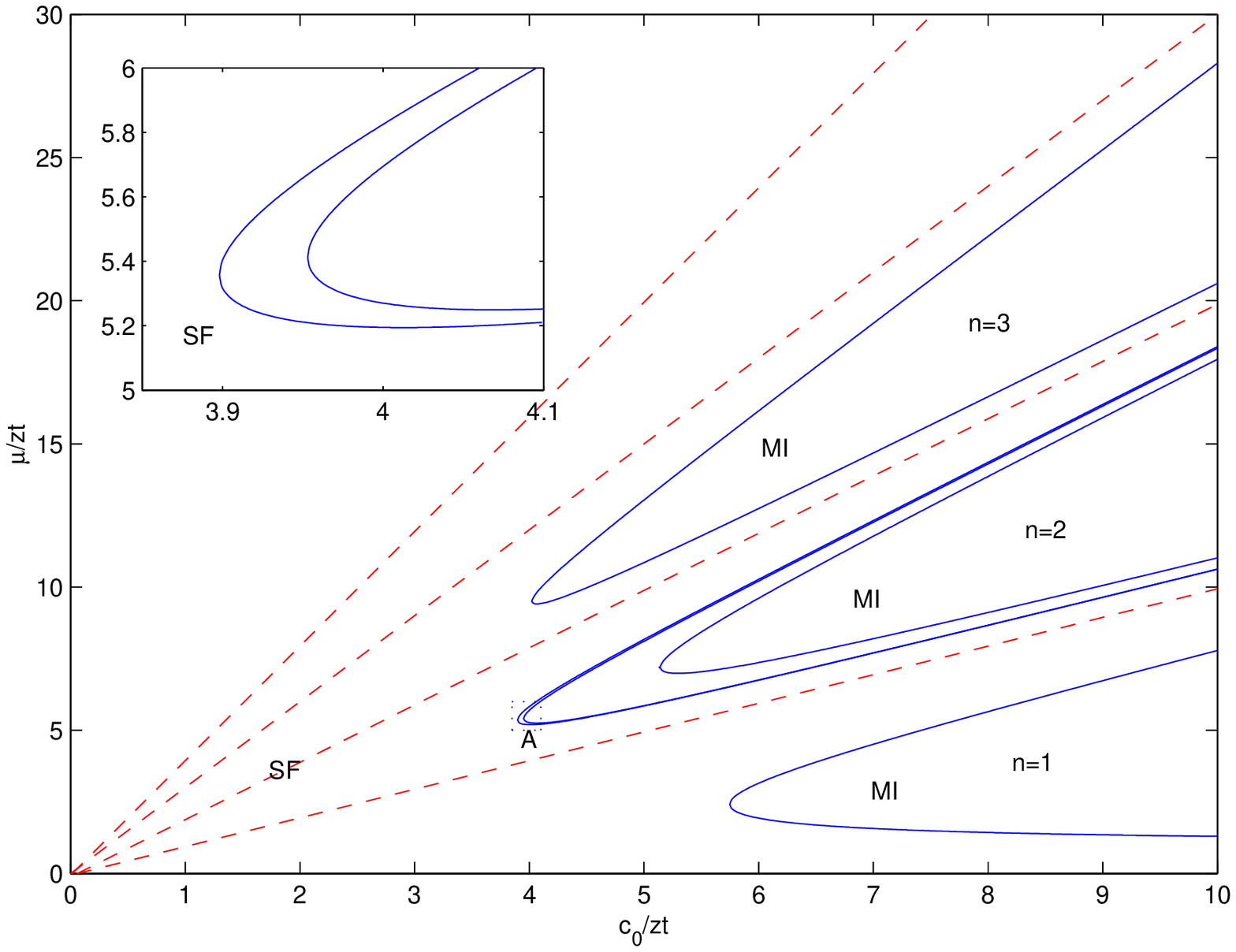} \caption{Phase diagram of
Bose-Hubbard Hamiltonian  obtained from second-order perturbation
theory with solid lines for $c_1=0.02zt$ and $c_2=0.1zt$. Here
$c_0/zt$ and $\mu/zt$ are dimensionless. The dashed lines indicate
the zeroth order phase diagram. In the diagram, SF and MI denote
superfluid phase and Mott insulator phase respectively. For $n=2$,
the interior line is the phase boundary of spin component with
Zeeman level $m=1$, and the external double lines, which are too
close to be distinguished in the diagram, are that of spin
components with $m=0,2$ respectively. The inset shows an expansion
of the part of the phase boundaries for $n=2$ in the dotted frame
labelled by A .} \label{2}
\end{figure}

\begin{figure}[ht]
\includegraphics[width=0.8\columnwidth]{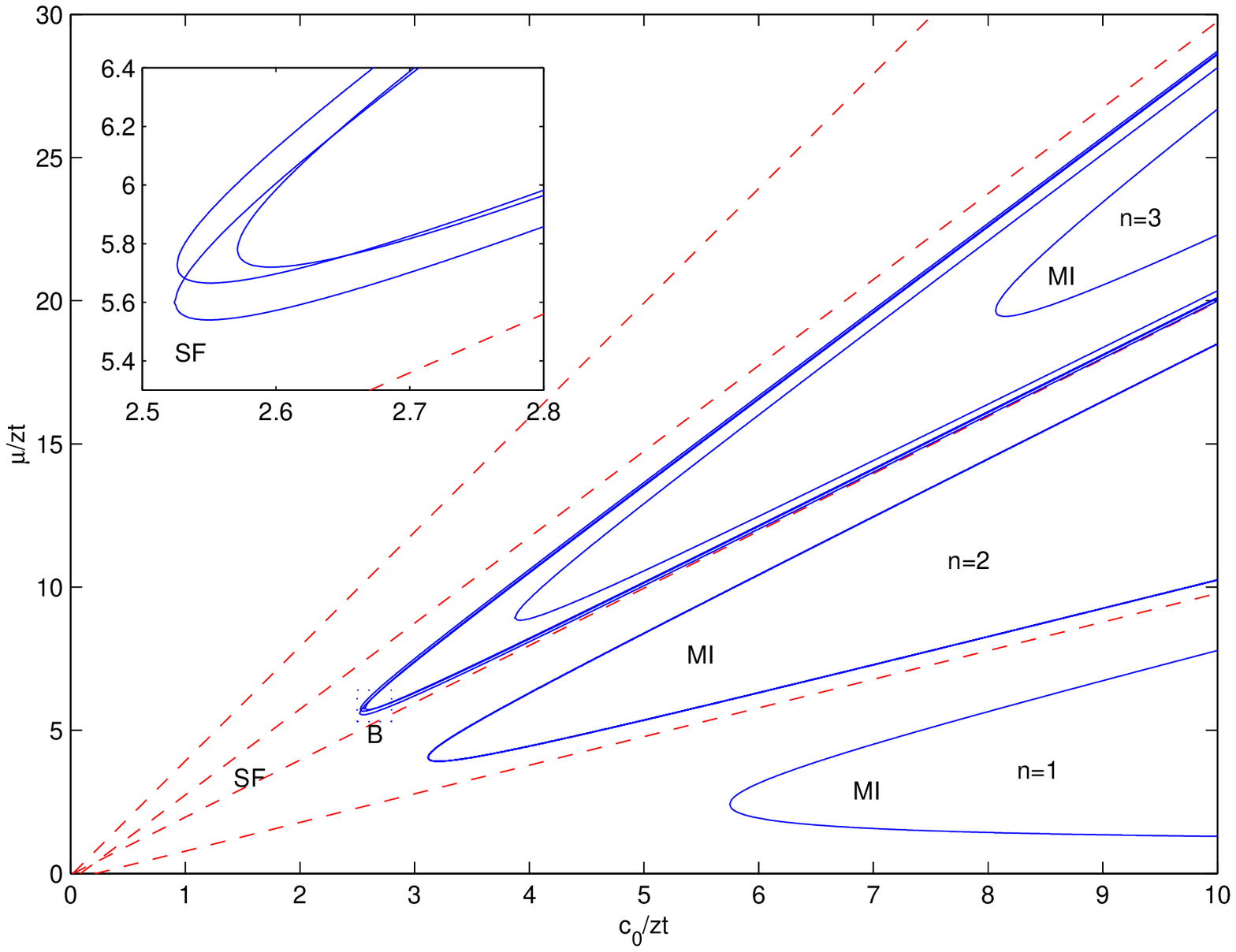}\caption{Phase diagram of
Bose-Hubbard Hamiltonian  obtained from second-order perturbation
theory with solid lines for $c_1=0.02zt$ and $c_2=-0.1zt$. Here
$c_0/zt$ and $\mu/zt$ are dimensionless. The dashed lines indicate
the zeroth order phase diagram. In the diagram, SF and MI denote
superfluid phase and Mott insulator phase respectively. For $n=3$
case, the interior line is the phase boundary of spin component
with Zeeman level $m=2$; the middle line the phase boundary of
spin component with Zeeman level $m=-2$, and the external triple
lines, which are too close to be distinguished in the diagram, the
phase boundaries of spin components with Zeeman level $m=0,\pm 1$.
The inset shows an expansion of the part of the phase boundaries
for $n=3$ in the dotted frame labelled by B.} \label{3}
\end{figure}

\begin{figure}[ht]
\includegraphics[width=0.8\columnwidth]{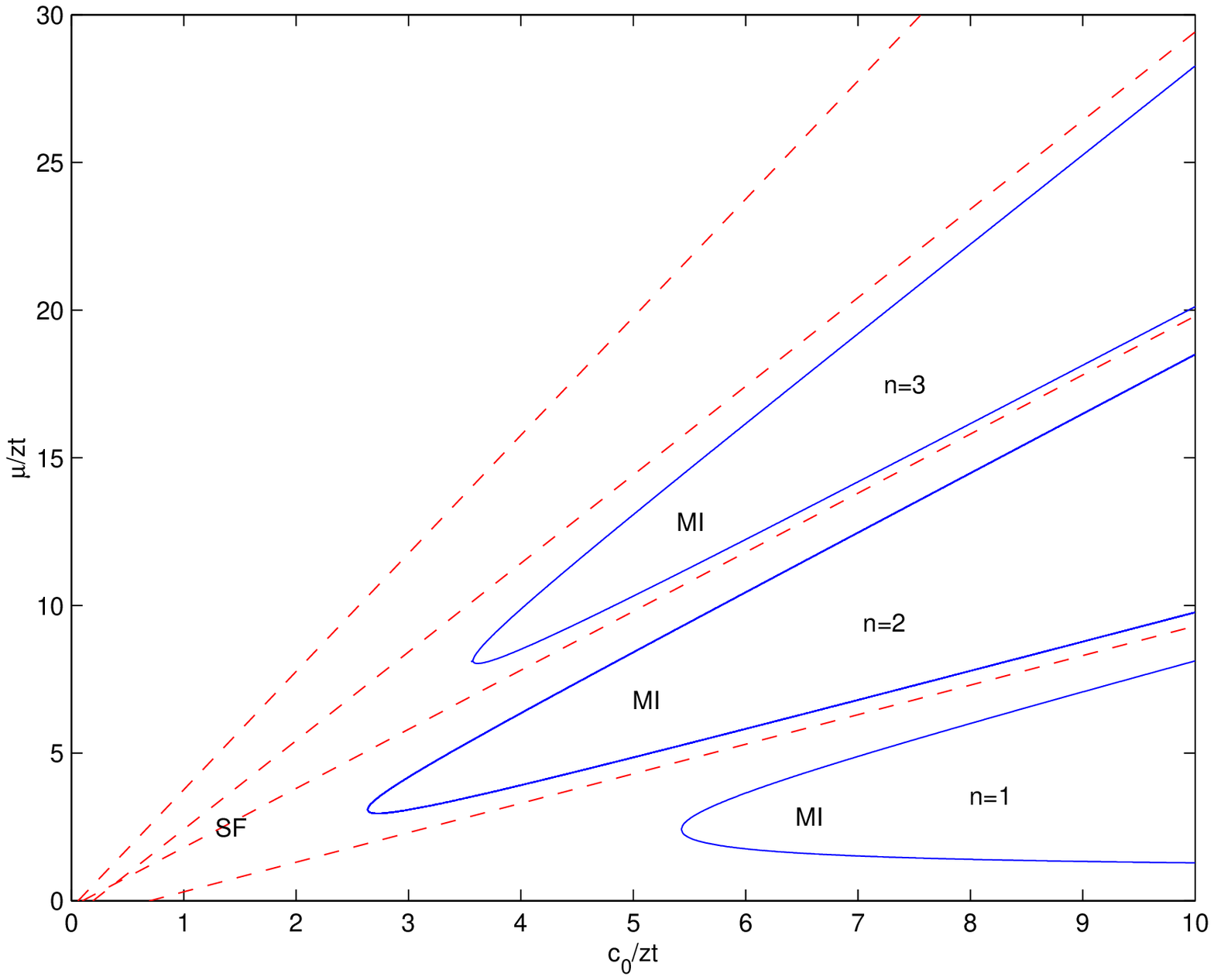} \caption{Phase diagram of
Bose-Hubbard Hamiltonian  obtained from second-order perturbation
theory with solid lines for $c_1=0.1zt$ and $c_2=-0.1zt$. Here
$c_0/zt$ and $\mu/zt$ are dimensionless. The dashed lines indicate
the zeroth order phase diagram. In the diagram, SF and MI denote
superfluid phase and Mott insulator phase respectively.} \label{4}
\end{figure}

\begin{figure}[ht]
\includegraphics[width=0.8\columnwidth]{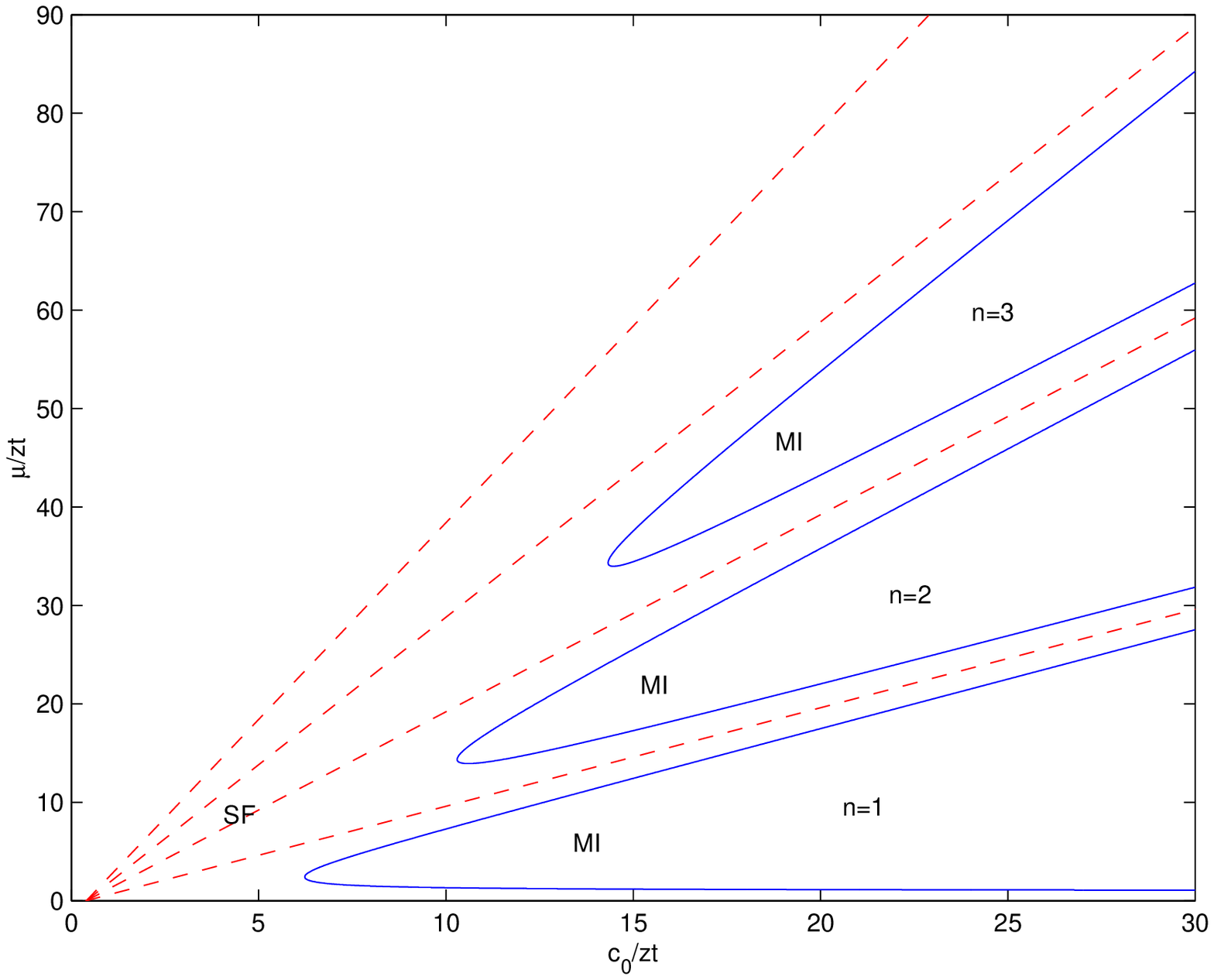}\caption{Phase diagram of
Bose-Hubbard Hamiltonian  obtained from second-order perturbation
theory with solid lines for $c_1=-0.1zt$ and $c_2=0.1zt$. Here
$c_0/zt$ and $\mu/zt$ are dimensionless. The dashed lines indicate
the zeroth order phase diagram. In the diagram, SF and MI denote
superfluid phase and Mott insulator phase respectively.} \label{5}
\end{figure}

\begin{figure}[ht]
\includegraphics[width=0.8\columnwidth]{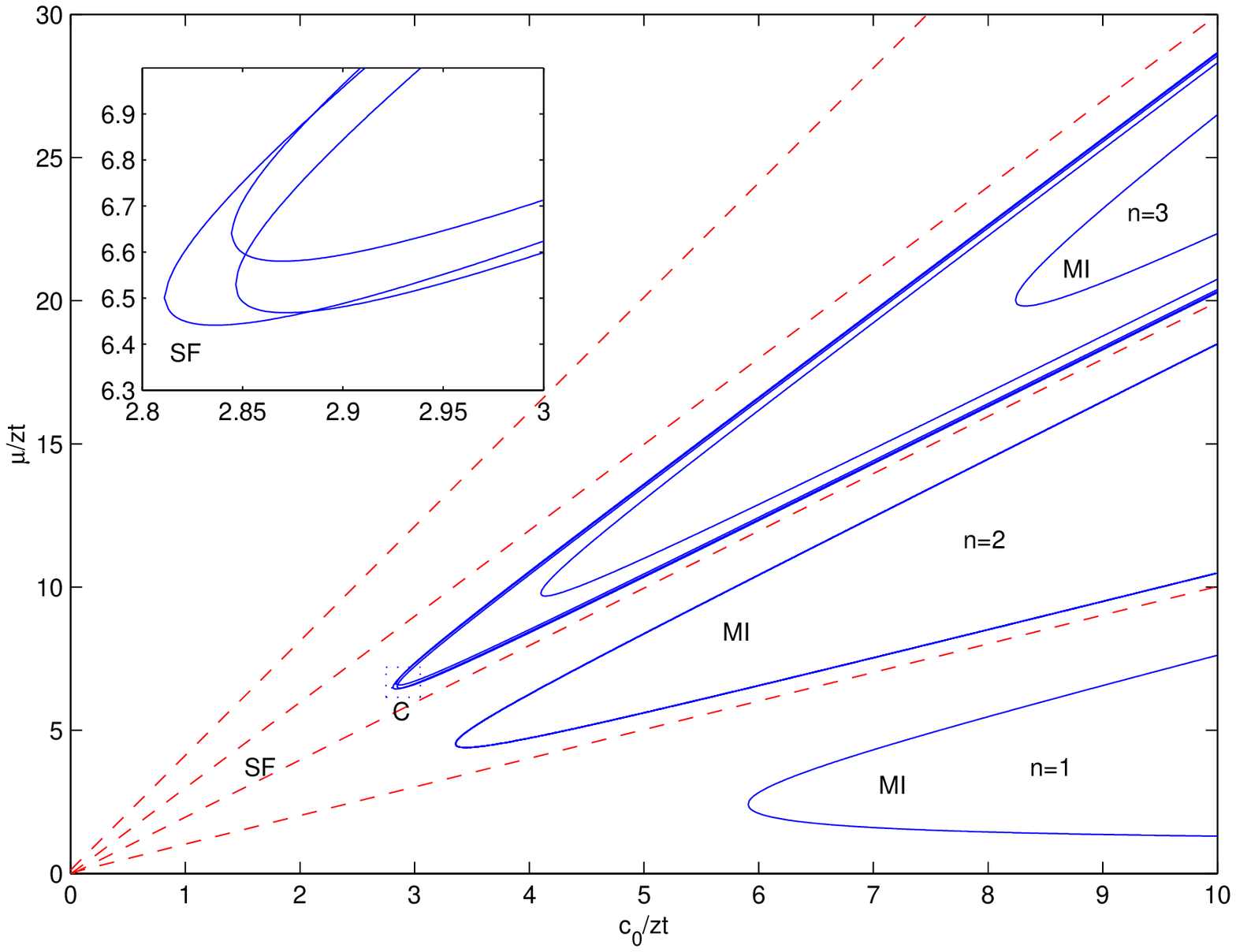}
\caption{Phase diagram of Bose-Hubbard Hamiltonian obtained from
second-order perturbation theory with solid lines for
$c_1=-0.02zt$ and $c_2=-0.1zt$. Here $c_0/zt$ and $\mu/zt$ are
dimensionless. The dashed lines indicate the zeroth order phase
diagram. In the diagram, SF and MI denote superfluid phase and
Mott insulator phase respectively. For $n=3$ case, the interior
line is the phase boundary of spin component with Zeeman level
$m=2$; the middle line the phase boundary of spin component with
Zeeman level $m=-2$, and the external triple lines, which are too
close to be distinguished in the diagram, the phase boundaries of
spin components with Zeeman level $m=0,\pm 1$. The inset shows an
expansion of the part of the phase boundaries for $n=3$ in the
dotted frame labelled by C.} \label{6}
\end{figure}

 Equiping the perturbative calculations we can draw  phase
diagrams. The phase diagrams show that  there is a phase
transition from Mott insulator with integer number bosons to
superfluid when the ratio $c_0/t$ is decreased to a critical
value. In the zeroth order, i.e., neglecting the hopping term, the
ground state is Mott state in which the occupation number per site
is pined at integer $n=1,2,...$, corresponding to a commensurate
filling of the lattice. Different Mott ground states maybe contain
different spin components. For example, there is only spin
component with Zeeman level $m=2$ when occupation number per site
$n=1$; spin components with $m=0,\pm 1,\pm 2$ for Mott state
$\prod_i(\hat{A}_0^{(2)}|0>)_i$ and spin components with $m=0,1,2$
for $\prod_i(\hat{A}_2^{(2)}|0>)_i$. It is easy to realize that
only one superfluid component appears when lowering the ratio
$c_0/t$ for the initial Mott ground state containing only one spin
component such as the case $n=1$ in Fig.\ref{1}-\ref{6} and
$n=1,2,3$ in Fig.\ref{5}. For the Mott ground states containing
multiple  spin components, when lowering the ratio $c_0/t$,
multiple superfluid components appear. The phase boundaries
between superfluid and Mott insulator phase for different spin
components are identical for some Mott ground states such as $n=2$
in Fig.1,3,4,6. and $n=3$ in Fig.1,2,4, and different for some
Mott ground states such as $n=2$ in Fig.2 and $n=3$ in Fig.3,6.

We find that  the position of phase boundary is related to average
occupation number of spin component in the initial Mott ground
state, i.e., the larger the average occupation number of spin
component per site is, the easier the transition from  Mott
insulator to superfluid phase. For example, in Fig.\ref{1}, the
average occupation numbers of spin component with Zeeman level
$m=2$ are 1, 2/5 and 3/5 per site for $n=1, 2, 3$ Mott ground
states respectively, and the minimum critical value of $c_0$ are
$5.42zt$, $2.76zt$ and $3.54zt$ for $n=1,2,3$ Mott states.  We can
arrive at the same result by analyzing  other figures.

\section{Conclusion}
In this paper, we have investigated the quantum phase transition
from Mott insulator to superfluid phase of spin-2 cold bosons at
zero temperature. First, we diagonalized   the Bose-Hubbard
Hamiltonian without  hopping term and got the eigenvalue and
ground states. Then taking the hopping term as a perturbation we
calculated  the first and second order corrections and drawn the
phase diagrams, which show that there is a phase transition from
Mott insulator with integer number bosons at each site to
superfluid phase when the ratio $c_0/t$ is decreased to a critical
value. Different Mott ground states maybe contain different spin
components. For some Mott ground states, the different spin
components  appear at different moment in superfluid phase when
lowering the ratio $c_0/t$.  The position of phase boundary is
related to average occupation number of spin component in the
initial Mott ground state.
\begin{acknowledgments} This work is in
part supported by NSF of China No.A0124015.
\end{acknowledgments}

\newpage

\begin{thebibliography}{99}
\bibitem{fisher}M.P.A.Fisher, P.B.Weichman, G.Grinstein, and
D.S.Fisher, Phys.Rev. B 40, 546(1989)
\bibitem{greiner}M.Greiner \textit{et al}., Nature 415,39(2002)
\bibitem{jaksch}D.Jaksch, C.Bruder, J.I.Cirac, C.W.Gardiner, and
P.Zoller, Phys.Rev.Lett. 81, 3108(1998)
\bibitem{oosten}D.van Oosten, P.van der Straten, and H.T.C.Stoof,
Phys. Rev. A 63, 053601(2001)
\bibitem{chen}G.H.Chen and Y.S.Wu
Phys. Rev. A 67, 013606(2003)
\bibitem{demler}E.Demler and F.Zhou, Phys. Rev. Lett. 88, 163001(2002)
\bibitem{tsuchiya}S.Tsuchiya,
S.Kurihara, and T.Kimura, cond-mat/0209676
\bibitem{stamper-kurn}D.M.Stamper-Kurn  \textit{et al}., Phys.Rev.Lett. 80,
2027(1998)
\bibitem{ho}T.L.Ho, Phys. Rev. Lett. 81, 742(1998)
\bibitem{ohmi}T.Ohmi and K.Machida, J. Phys. Soc. Jpn. 67,
1822(1998)
\bibitem{ciobanu}C.V.Ciobanu, S.K.Yip and T.L.Ho, Phys. Rev.
A 61, 033607(2000)
\bibitem{koashi}M.Koashi and M.Ueda, Phys. Rev. Lett. 84, 1066(2000)
\bibitem{koashi2}M.Ueda and M.Koashi, Phys. Rev. A 65,
063602(2002)
\end{thebibliography}
\end{document}